\documentclass[submission, Phys]{SciPost}
\usepackage{graphicx}
\usepackage{amsmath}
\usepackage[square,sort,compress,numbers]{natbib}

\newcommand{\ket}[1]{\left|{#1}\right>}

\def\pdagger{{\phantom{\dagger}}}

\newcommand{\D}{\mathrm{d}}
\newcommand{\C}{\mathrm{c}}
\newcommand{\rsS}{{\rm \scriptscriptstyle S}}
\newcommand{\rsD}{{\rm \scriptscriptstyle D}}

\newcommand\GS{{\it{\Gamma}_{\rm{S}}}}
\newcommand\GD{{\it{\Gamma}_{\rm{D}}}}

\newcommand\Vcav{V_{\rm c}}
\newcommand\Vdot{V_{\rm d}}
\newcommand\V{V_{\rm SD}}

\newcommand\eV{{\rm eV}}

\newcommand{\rs}{\rm \scriptscriptstyle}
\newcommand{\s}{\scriptscriptstyle}
\newcommand{\rsL}{\rm \scriptscriptstyle S}
\newcommand{\rsR}{\rm \scriptscriptstyle D}

\begin{document}

\begin{center}{\Large \textbf{
Transport spectroscopy of singlet-triplet quantum dot states coupled to electronic cavities
}}\end{center}

\author{}
\begin{center}
M. S. Ferguson\textsuperscript{1*},
C. R{\"o}ssler\textsuperscript{2},
T. Ihn\textsuperscript{3},
K. Ensslin\textsuperscript{3},
G. Blatter\textsuperscript{3},
O. Zilberberg\textsuperscript{3}
\end{center}

\begin{center}
{\bf 1} Institute for Theoretical Physics, ETH Zurich, 8093 Z{\"u}rich, Switzerland
\\
{\bf 3} Infineon Technologies Austria, Siemensstra{\ss}e 2, 9500 Villach, Austria
\\
{\bf 2} Solid State Physics Laboratory, ETH Zurich, 8093 Z{\"u}rich, Switzerland
\\
* femichae@phys.ethz.ch
\end{center}

\begin{center}
\today
\end{center}


\section*{Abstract}
{\bf
A strong coupling between an electronic cavity and a quantum dot has been recently demonstrated [Phys. Rev. Lett. 115, 166603 (2015)]  and described in a comprehensive theoretical framework [Phys. Rev. B 96, 235431 (2017)]. Here, we focus on the signatures that demonstrate the cavity's impact on inelastic singlet-triplet transport through the dot. We find the same transport signatures in the experiment as predicted by the model that describes the coupled dot--cavity system. Interestingly, a lowest-order treatement of the coupling to the electronic leads on top of an exact diagonalisation of the dot--cavity system is sufficient to highlight the interplay between the cavity and the higher-order inelastic singlet-triplet cotunneling.
}


\section{Introduction}

Confinement of two dimensional electronic gases in ultraclean materials have been used with great success to create a myriad of quantum devices~\cite{ihn2010semiconductor}. Transport through these devices  displays fascinating low-dimensional coherent phenomena at mesoscopic scales. Having established a firm control over the fundamental building blocks of the mesoscopic heterostructures, nowadays, various mesoscopic structures are successfully coupled to each other, leading to new phenomena and applications. For example, in a recent work~\cite{roessler:2015,ferguson_2017}, we were able to demonstrate spin-coherent coupling between an electronic cavity~\cite{katine_point_1997,hersch_diffractive_1999} and a few-electron quantum dot~\cite{kouwenhoven_few-electron_2001}. 

The device offers a tunable method for this coupling, exhibiting a spin-singlet state that extends over the entire device, on length scales of $\sim$2$\mu$m, and generates cavity-assisted cotunneling processes.
We report, here, on similar transport spectroscopy measurements of the dot--cavity system as were reported in Ref.~\cite{roessler:2015}. In the previous work, we focused on the effect of the electronic cavity on Kondo transport, i.e., when the dot has an odd number of electrons. Here, we focus on a regime where the dot is populated by an even number of electrons. We experimentally observe, once more, cavity-assisted cotunneling processes, but also highlight the amplification of inelastic singlet-triplet cotunneling~\cite{furusaki1995theory,zumbuhl2004cotunneling} through the dot by the cavity. Following Ref.~\cite{ferguson_2017} we write an effective model, which we analyze using a rate equation method, thus, predicting the out of equilibrium transport through the system corresponding to the exerimental data.

\section{Dot--cavity experiment}
\begin{figure}[h]
	\centering
	\includegraphics[width=14pc]{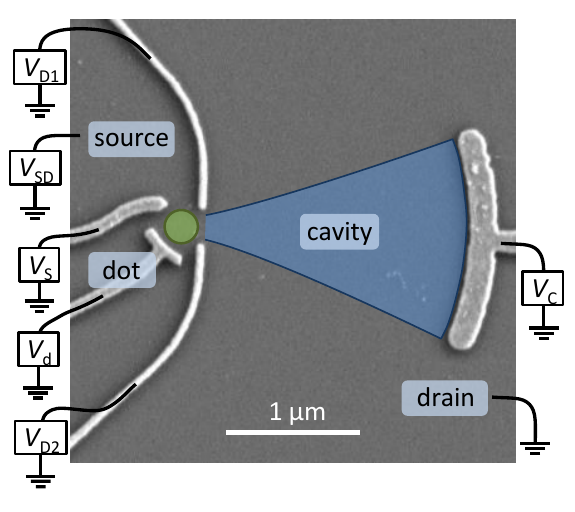}\hspace{2pc}%
	\begin{minipage}[b]{14pc}\caption{\label{Fig1} Scanning electron micrograph of the dot--cavity device, see also Ref.~\cite{roessler:2015}. Schottky
			electrodes (bright) define an electronic cavity (Blue overlay) focused onto a
			dot (green circle overlay). Gate voltages $V_{\rm S}$ and $V_{\rm Di}$ with $i=1,2$ control the dot tunnel coupling to source and drain, respectively. Gate voltages $V_{\rm d}$ and $V_{\rm C}$ control the energies and occupancies of the dot and cavity, respectively.}
	\end{minipage}
\end{figure}

The electronic dot--cavity device is shown in Fig.~\ref{Fig1}.
A two-dimensional electron gas (2DEG) is formed $90\,\rm{nm}$ underneath the surface of a GaAs/AlGaAs
heterostructure. Applying negative voltages to Schottky top gates depletes the
underlying 2DEG and defines a quantum dot that is tunnel-coupled to source and drain
leads. An additional curved gate is positioned $\sim 2\,{\rm \mu m}$ away
from the dot. Applying a voltage $\Vcav$ generates an electronic mirror that confines
quantized ballistic cavity modes with increased weight at the tunnel
barrier~\cite{hersch_diffractive_1999,ferguson_2017}. The mirror gate
has a relatively small opening angle of $45^\circ$ in order to confine only
fundamental one-dimensional modes, i.e., high angular-momentum modes leak
out into the drain.

The specific device design allows for controllable transport spectroscopy of a few-electron dot, i.e., provides enough degrees of freedom to tune relevant experimental parameters independently. The gate voltage $V_{\rm D1}=-0.4\,\rm{V}$ controls the position of the dot potential minimum between the source and drain tunnel barriers. It was kept unchanged throughout the experiments. The dot plunger gate voltage $V_{\rm d}\sim-0.65\,\rm{V}$ changes the occupation number of the dot. The tunnel barrier gates $V_{\rm S}$ and $V_{\rm D2}$ control the tunnel couplings, $\GS$ and $\GD$, of the dot to the source and drain leads, respectively.

\begin{figure}
	\begin{center}
		\includegraphics{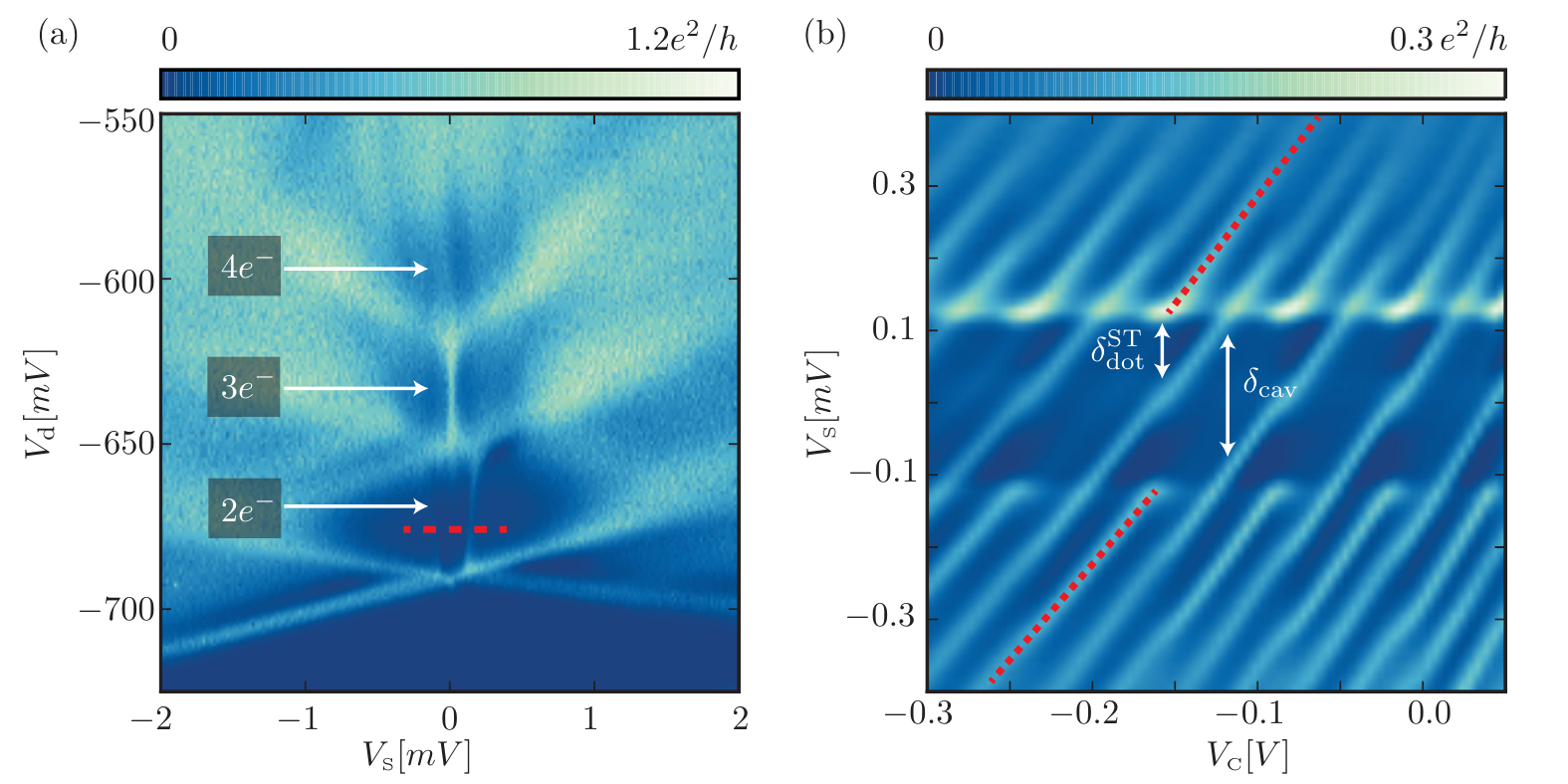}
	\end{center}
	\caption{\label{Fig2}Finite-bias measurements of the differential conductance $g=dI/d\V$ through the dot--cavity device. The tunnel
		couplings were tuned to be symmetric $\GD\approx \GS$. (a) A standard Coulomb diamond structure as a function of $\V$ and
		$\Vdot$ is observed when the cavity is switched off ($\Vcav=+200\,\rm{mV}$). Characteristic (dark) regions of Coulomb blockade are intersected by a pronounced zero-bias Kondo resonance at odd dot occupation ($N=3\,e^-$). At even occupation ($N=2\,e^-$), a ubiquitous singlet-triplet inelastic feature is seen at finite bias. 
		(b) Same experimental parameters as highlighted by the red dashed line in (a), but as a function of $\Vcav$. Additional resonance lines spaced by $\delta_{\rm cav}\approx~220\,\mu \eV$ arise due to the cavity modes modifying the dot transport. The continuous lines are due to elastic cavity assisted cotunneling processes while the discontinuous lines split off by \(\delta_\mathrm{dot}^\mathrm{ST}\) (one set indicated by red dotted lines) are due to inelastic cavity assisted cotunneling processes and thus only appear at finite bias. Please note that the ratio \(\delta_\mathrm{cav}/\delta_\mathrm{dot}^\mathrm{ST}\approx 2\) is not constrained, and may be different for different experimental parameters.} 
\end{figure}

We focus, here, on a subset of transport spectroscopy measurements through the dot--cavity system (cf.~Ref.~\cite{roessler:2015}). The experiments were conducted at an electronic temperature $T_{\rm{el}}<20\,\rm{mK}$~\cite{baer_experimental_2014, rossler_spectroscopy_2014}. In Fig.~\ref{Fig2}, we report on two finite-bias measurements of the differential conductance $g=dI/d\V$ through the dot--cavity device. In Fig.~\ref{Fig2}(a), a standard Coulomb diamond structure is observed when the cavity mirror gate is switched off ($\Vcav=+200\,\rm{mV}$), cf.~Fig.~2(a) in Ref.~\cite{roessler:2015}. We observe a pronounced Kondo--resonance in the $N=3\,e^-$ charge state. Important for our discussion here, is the appearance of a singlet-triplet inelastic cotunneling feature within the $N=2\,e^-$ Coulomb blockade valley. Additionally, such a measurement allows us to estimate the model parameters of our system, for example, the tunnel coupling constants are estimated by inspection of the pronounced Kondo--resonance in the $N=3\,e^-$ charge state. From the full-width-half-maximum of the Kondo--resonance we estimate a Kondo--temperature of $T_{\rm K}\approx100\,\rm{mK}$. After determining the dot's charging energy $U\approx700\,\rm{\mu eV}$ from the extent of the Coulomb diamond, we use the relation: $T_{\rm K}=1/2\sqrt{{\it \Gamma} U}{\rm e}^{-\pi U/4{\it \Gamma}}$ and obtain ${\it \Gamma}_{\rm S}={\it \Gamma}_{\rm D}={\it \Gamma}/2\approx87\,\rm{\mu eV}$. Similarly, from the singlet-triplet inelastic cotunneling feature, we observe the singlet-triplet splitting on the dot ${\it \delta}_{\rm dot}^{\mathrm{ST}}\approx 110\,\rm{\mu eV}$.

Positioning the dot deep within this valley, the cavity gate is switched on gradually by applying bias on $\Vcav$, see Fig.~\ref{Fig2}(b) (cf.~Fig.~4(f) in Ref.~\cite{roessler:2015}). Once the 2DEG below the cavity gate is depleted, an electronic cavity is formed with its states filled up to the chemical
potential. Considering the lithographically defined distance of the mirror gate from the drain tunnel barrier of the dot $L_{\rm cav}=1.9\,{\rm \mu m}$ and a Fermi wavelength $\lambda_{\rm F}\approx 53\,{\rm nm}$, we estimate that upon formation $n_{\rm cav}\approx 2L_{\rm cav}/\lambda_{\rm F}\approx 70$ states are filled.  Applying increasingly
negative $\Vcav$, the mirror gate depletes more of the 2DEG below it and makes the cavity shorter. Thus, the cavity level-spacing becomes larger and its states rise in energy, passing through the chemical potential, and effectively enhancing the tunnel
coupling $\GD$ by doing so. This is seen in Fig.~\ref{Fig2}(b) as a series of peaks with enhanced cavity-assisted cotunneling. 

\section{Dot--cavity theory}
In the following, we introduce an effective model describing the dot-cavity system~\cite{oehri_prep_2016}. This corresponds to the Hamiltonian
\begin{align}\label{eq:ham}
H=H_{\mathrm{leads}}+H_{\mathrm{dot}}+H_{\rm cav}
+H_{\rm coupl}+H_{\rm tun}\,,
\end{align}
where 
\begin{align}
H_{\mathrm{leads}}
=\sum_{s,\sigma} \epsilon_{s} c_{s\sigma}^\dagger c_{s\sigma}^\pdagger
+\sum_{d,\sigma} \epsilon_{d} c_{d\sigma}^\dagger c_{d\sigma}^\pdagger\,,
\end{align}
describes the source and drain leads with creation and annihilation operators of lead states $c_{i\sigma}^\dagger$ and $c_{i\sigma}^\pdagger$, with spin \(\sigma\) and $i=s,d$ indicating momenta of states in the leads, respectively. For our discussion, it is sufficient to consider a dot described by two interacting levels
\begin{align}
H_{\mathrm{dot}}
=\sum_{\sigma}\Big[\epsilon_{\D} d_{1\sigma}^\dagger d_{1\sigma}^\pdagger + \left(\epsilon_{\D} + {\it \delta}_{\rm d}\right) d_{2\sigma}^\dagger d_{2\sigma}^\pdagger\Big]
+Un(n-1)/2\,,
\label{Eq:dotH}
\end{align}
with creation and annihilation operators $d_{j\sigma}^\dagger$ and $d_{j\sigma}^\pdagger$ of the dot levels, $j=1,2$ with energies $\epsilon_\D$ and $\left(\epsilon_{\D} + {\it \delta}_{\rm d}\right)$, and spin $\sigma$. Additionally, the occupation operator is $n=\sum_{j,\sigma} d^\dagger_{j\sigma}d^\pdagger_{j\sigma}$, and Coulomb interaction is denoted by $U$. The cavity Hamiltonian describes a sum of discrete non-interacting levels
\begin{align}
H_{\mathrm{cav}}=\sum_{\sigma,m} (\epsilon_{\C}+m\delta_{\C})
f_{m\sigma}^\dagger f_{m\sigma}^\pdagger\,,
\label{Eq:cav}
\end{align}
with cavity energy $\epsilon_{\C}$, equal cavity spacing \(\delta_{\C}\), integer \(m\) and creation and annihilation operators $f_{m\sigma}^\dagger$ and $f_{m\sigma}^\pdagger$. The dot and cavity coupling Hamiltonian is
\begin{align}
H_{\mathrm{coupl}}&=\sum_{j,m,\sigma} \Omega_{jm}  f_{m\sigma}^\dagger 
d_{j\sigma}^\pdagger
+\mathrm{h.c.}\,,
\label{Eq:coupl}
\end{align}
with tunneling amplitudes $\Omega_{jm}$, and the tunneling Hamiltonian between the dot-cavity central area and the leads is
\begin{align}
H_{\mathrm{tun}}&= 
H_{\mathrm{tun}}^{\D\rsL}+H_{\mathrm{tun}}^{\D\rsR}+H_{\mathrm{tun}}^{\C\rsR}
\nonumber\\
&=t_{\rsL}\sum_{s, j,\sigma} ( d_{j\sigma}^\dagger 
c_{s\sigma}^\pdagger +\mathrm{h.c.})
+t_{\rsR}\sum_{d, j,\sigma} ( c_{d\sigma}^\dagger
d_{j\sigma}^\pdagger +\mathrm{h.c.})+t_\C\sum_{m,s,\sigma} ( c_{s\sigma}^\dagger 
f_{m\sigma}^\pdagger+\mathrm{h.c.})\,.
\label{eq:Htun}
\end{align}
The tunneling amplitudes $t_{\rm a}$ couple the dot to source ($\mathrm{a}=\mathrm{S}$) and drain leads ($\mathrm{a}=\mathrm{D}$), and correspond to the rates $\Gamma_{\rm a}=2\pi \rho_{\rm a} \left|t_{\rm a}\right|^2$. The amplitude $t_\C$ describes the coupling between cavity levels and the drain lead (which we assume is the same for all cavity levels), and gives rise to the rate, $\Gamma_{\C}=2\pi \rho_{\rsR} \left|t_{\C}\right|^2$. Here, and in the following, we set $\hbar=1$.

\subsection{Master equation analysis}
\begin{figure}
	\begin{center}
		\includegraphics{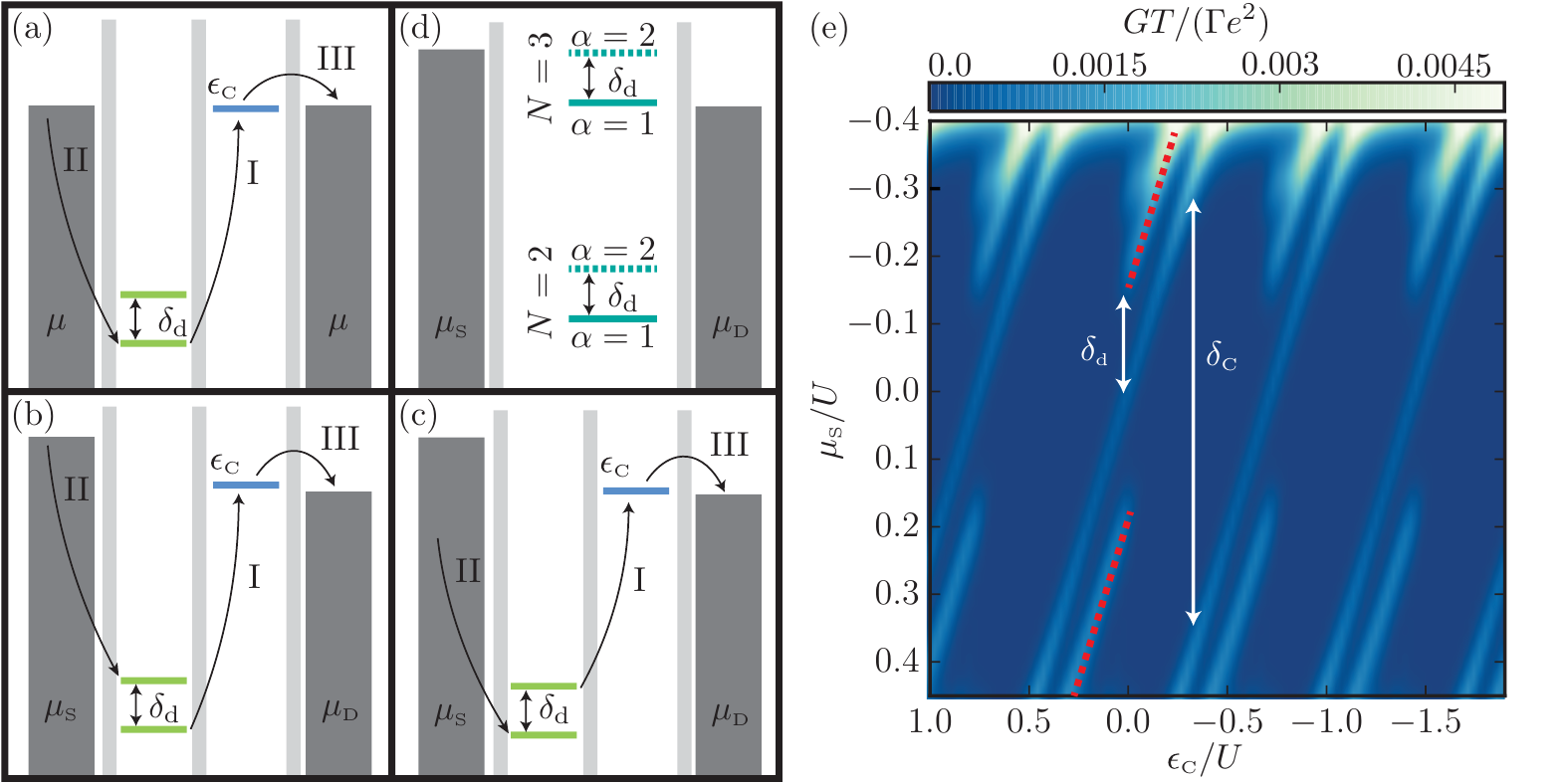}
	\end{center}
	\caption{\label{Fig3} Examples of relevant processes and results deep within an even occupation Coulomb blockade valley. (a) An elastic equilibrium cavity-assisted cotunelling process where an electron virtually hops out of the dot into the cavity (I),  and is replaced by an electron from the source (II). Finally, the electron in the cavity tunnels resonantly into the drain (III). (b) An inelastic cavity-assisted cotunneling process. As in (a), an electron first virtually hops into the cavity (I) but then the dot is populated in its triplet excited state by a source electron (II). Here, too, the electron in the cavity tunnels resonantly into the drain (III). (c) Similar to (b), but here the process takes an excited triplet state into the singlet ground-state. The processes (a)-(c), as well as, additional ones that are not illustrated, are automatically taken into account by our master equation as the central region is exactly diagonalized. Indeed, in our sequential transport approach only processes including two or more lead operators are neglected: a simple illustration of this can be seen in (d) where the exactly diagonalized dot--cavity hybrid system is depicted in the addition spectrum representation. An excited (\(\alpha > 1\)) state can only be probed if the absolute value of the difference in chemical potentials \(\Delta \mu=\mu_{\rsD}-\mu_{\rsL}\) is larger than the splitting \(\delta=\epsilon^{\alpha}-\epsilon^{1}\). (e) The calculated differential conductance using the master equation approach [cf.~Eq.~\eqref{eq:master_current}]. The parameters used are \(\mu_{\rsD}=0\), \(T/U=0.02\), \(\Omega_{jm}/U=0.05\), \(\delta_{\C}/U=0.75\), \(\delta_{\D}/U=0.2\) and \( \epsilon_{\D}=-1.5 U+0.5 \mu_{\rsS} +0.1 \epsilon_{\C} \). At low bias, a continuous cavity-assisted cotunneling line appears corresponding to process (a). At sufficiently large bias \(|\mu_{\rsS}|\geq \delta_\D\), discontinuous lines (one set indicated by red dotted lines) appear due to cavity-assisted inelastic cotunneling transport through the excited triplet state, see also (b), (c), and (d). Here we capture the experimental signatures, and can confirm that the model parameters \(\delta_{\D}\) and \(\delta_{\C}\) correspond to the experimentaly established parameters \(\delta_\mathrm{dot}^\mathrm{ST}\) and \(\delta_\mathrm{cav}\) from Fig.~\ref{Fig2}(b) respectively.}
\end{figure}

We employ a master equation approach to describe the transport through the dot--cavity interacting system, assuming that it is weakly-coupled to the leads~\cite{sakurai1985, beenakker:91, Korotkov:1994, Koch:2006,ferguson_2017}. We are interested in describing the effect of the cavity on inelastic-cotunneling processes through the dot, see Fig.~\ref{Fig3}(a)-(c) for illustrated examples. Here, we do not analyze high-order cotunneling rates through the dot. As we shall see, to capture the effect, it is sufficient to consider sequential transport through the exact solutions of the central (closed) dot--cavity system, i.e., by making use of the exact diagonalization results, we consider the dot--cavity coupling to all orders, but consider the couplings to the leads in lowest order, see Fig.~\ref{Fig3}(d).

Let us first consider, the dot--cavity Hamiltonian [cf.~Eqs.~\eqref{Eq:dotH}, \eqref{Eq:cav}, and \eqref{Eq:coupl}] 
\begin{align} 
\label{eq:molecule} 
H_{\rm dc}=H_{\rm dot}+H_{\rm cav}+H_{\rm coupl}\,. 
\end{align} 
Assuming that only $n$ cavity levels are close to the chemical potential and affect the physics of the device, we can solve this Hamiltonian by exact diagonalization of a $2^{2n+4}\times 2^{2n+4}$ matrix, and determine the eigenstates and eigenenergies of the isolated dot--cavity system. 
The obtained set of solutions, $\epsilon_{N}^{\alpha}$, $\ket{N,\alpha}$ can be characterized by the occupation (electron number, $N$) of the central region (Fock space), and by sorting the eigenenergies $\epsilon_{N}^{\alpha}$ within each Fock block in increasing order and labeling them by $\alpha$, i.e., $\alpha=1$ denotes the ground state with $N$ electrons.

The occupation probabilities $P^{\alpha}_{N}$ of populating the eigenstates when the system is coupled to leads are found by determining the steady state of the coupled rate equations (master equation)
\begin{align}
\partial_{t} P^{\alpha}_{N}=
\sum_{N',\alpha'} (W^{\alpha,\alpha'}_{N,N'} 
P^{\alpha'}_{N'}
-W^{\alpha',\alpha}_{N',N} P^{\alpha}_{N})\,,
\end{align}
where the rate $W^{\alpha,\alpha'}_{N,N'}$ corresponds to the transition rate from state $\ket{N',\alpha'}$ to state $\ket{N,\alpha}$. In the following, we only consider coupling to the leads, $H_{\mathrm{tun}}$,  to lowest order (sequential tunneling), i.e., we consider only the rates $W_{(N\pm 1),N}$. The sequential tunneling rates are composed of three contributions,
\begin{align}
W_{(N\pm 1),N}
=
W^{\s \D \rsL}_{(N\pm 1),N}
+W^{\s \D \rsR}_{(N\pm 1),N}
+W^{\s \C \rsR}_{(N\pm 1),N}\,,
\end{align}
and are given by
\begin{align}
W^{\s \D \rs a}_{(N\pm 1),N}
&=\Gamma_{\rs a} | \langle \psi_{N\pm 1} |
\hat{c}_{d\sigma}^\pm| \psi_{N}\rangle|^2 g^{\rs a}_{\pm}(\epsilon_{N\pm 1}-\epsilon_{N})\,,\label{eq:rate1}\\
W^{\s \C \rsR}_{(N\pm 1),N}
&=\Gamma_{\C} | \langle \psi_{N\pm 1} |
\sum_{j}\hat{f}_{j\sigma}^\pm| \psi_{N}\rangle|^2 g^{\rs R}_{\pm}(\epsilon_{N\pm 1}-\epsilon_{N})\,,\label{eq:rate2}
\end{align}
where $g^{\rs a}_{+}(\epsilon)=n_{\rs F}(\epsilon-\mu_{\rs a})$ and $g^{\rs a}_{-}(\epsilon)=1-n_{\rs F}(-\epsilon-\mu_{\rs a})$ are, respectively, the electron and hole distribution functions with $n_{\rs F}(\epsilon)=1/(1+e^{\beta \epsilon})$  the Fermi-Dirac function, and $\beta=1/k_{\rs B}T$. We use the operator notation  $\mathcal{O}^{+}=\mathcal{O}^\dagger$ and $\mathcal{O}^{-}=\mathcal{O}$.  

The current through the system is expressed through the rates and the steady state occupation probabilities as
\begin{align}
I=e \sum_{N,\sigma}
(W_{(N+1),N}^{\D \rm L}-W_{(N-1),\mathrm{N}}^{\D\rm L})
P_{N}\,.
\label{eq:master_current}
\end{align}

In Fig.~\ref{Fig3}(e), we plot the calculated differential conductance $G=dI/d\V$. Our model predicts the (elastic) cavity-assisted cotunneling signatures in the low bias regimes, highlighting the contribution of processes such as the one depicted in Fig.~\ref{Fig3}(a). Additionally, at higher-bias, the signatures of inelastic singlet-triplet cotunneling appear as resonant processes of the hybrid dot--cavity system, cf.~Figs.~\ref{Fig3}(b) and (c). These features are captured within our sequential tunneling approach due to the fact that we have taken the dot--cavity coupling to all orders, cf.~Fig.~\ref{Fig3}(d). Missing signatures in our treatment include the direct inelastic cotunneling through the dot when the cavity is non-resonant, cf.~horizontal lines in Fig.~\ref{Fig2}(b). Please note, that we work in a regime where temperature dominates the width for the various energy levels, $T \gg \GS, \GD$.

\section{Conclusion}
We have shown that strong coupling between an electronic cavity and a quantum dot leads to cavity-assited cotunneling. Specifically, when the dot is tuned deep within an even occupation valley, clear signatures of cavity resonances appear. The cavity also directly affects signatures of inelastic singlet-triplet cotunneling in this device. We have developed a theoretical model where we observe that such higher-order inelastic cotunneling can be treated to lowest-order in the coupling to leads via a resonant cavity mode. Such tunable amplification can be useful for cavity-assisted spectroscopy of excited states of a quantum dot.

\section{Acknowledgments}

\paragraph{Funding information} We acknowledge financial support from the Swiss National Science Foundation, Division 2, and through the National Centre of Competence in Research “QSIT - Quantum Science and Technology”.

\nolinenumbers

\end{document}